\documentclass[preprintnumbers,amsmath,amssymbm,preprint]{revtex4}
\usepackage{epsfig}
\usepackage{graphicx}
\usepackage{amssymb}

\begin{document}

\title{Lower bound on the radii of circular orbits in the extremal Kerr black-hole spacetime}
\author{Shahar Hod}
\affiliation{The Ruppin Academic Center, Emeq Hefer 40250, Israel}
\affiliation{ } \affiliation{The Hadassah Institute, Jerusalem
91010, Israel}
\date{\today}

\begin{abstract}

\ \ \ It is often stated in the physics literature that
maximally-spinning Kerr black-hole spacetimes are characterized by
near-horizon co-rotating circular geodesics of radius
$r_{\text{circular}}$ with the property $r_{\text{circular}}\to
r^+_{\text{H}}$, where $r_{\text{H}}$ is the horizon radius of the
extremal black hole. Based on the famous Thorne hoop conjecture, in
the present compact paper we provide evidence for the existence of a
non-trivial lower bound
${{r_{\text{circular}}-r_{\text{H}}}\over{r_{\text{H}}}}\gtrsim
(\mu/M)^{1/2}$ on the radii of circular orbits in the extremal Kerr
black-hole spacetime, where $\mu/M$ is the dimensionless mass ratio
which characterizes the composed black-hole-orbiting-particle
system.
\end{abstract}
\bigskip
\maketitle

\section{Introduction}

The circular geodesic motions of test particles in rotating Kerr
black-hole spacetimes have attracted the attention of physicists and
mathematicians during the last five decades (see
\cite{tw1,tw2,tw3,tw4} and references therein). The characteristic
radii of these astrophysically important orbits are bounded from
below by the radius of the equatorial null circular geodesic which,
for a given value of the black-hole angular momentum, is
characterized by the smallest possible circumference.

Intriguingly, as discussed by many authors
\cite{tw1,tw2,tw3,tw4,Ted1,Ted2,Ted3}, in the maximally-spinning
(extremal) Kerr black-hole spacetime, there exist circular orbits of
radius $r_{\text{circular}}$ which are characterized by the limiting
near-horizon behavior $r_{\text{circular}}/r_{\text{H}}\to 1^+$
\cite{Noteext,Noteunit}.

The main goal of the present compact paper is to provide evidence,
which is based on the famous Thorne hoop conjecture \cite{Thorne},
for the existence of a non-trivial lower bound
$r_{\text{circular}}>r_{\text{min}}>r_{\text{H}}$ on the radii of
circular orbits in the maximally-spinning Kerr black-hole spacetime.
In particular, as we shall show below, according to the hoop
conjecture the smallest possible circular radius
$r_{\text{min}}=r_{\text{min}}(\mu/M)$ is determined by the
dimensionless mass ratio $\mu/M$ \cite{Notemu} which characterizes
the composed extremal-black-hole-orbiting-particle system.

\section{Circular orbits in the maximally-spinning Kerr black-hole spacetime}

We consider a particle of proper mass $\mu$ which orbits around a
maximally-spinning (extremal) Kerr black-hole of mass $M$ (with
$\mu/M\ll1$) and angular momentum $J=M^2$. The extremal curved
black-hole spacetime is characterized by the line element \cite{tw1}
\begin{eqnarray}\label{Eq1}
ds^2=-(1-2Mr/\Sigma)dt^2-(4M^2r\sin^2\theta/\Sigma)dt d\phi
+(\Sigma/\Delta)dr^2+\Sigma
d\theta^2 \nonumber \\ +(r^2+M^2+2M^3r\sin^2\theta/\Sigma)\sin^2\theta d\phi^2\ ,
\end{eqnarray}
where $(t,r,\theta,\phi)$ are the familiar Boyer-Lindquist spacetime
coordinates and the metric functions in (\ref{Eq1}) are defined by
the relations
\begin{equation}\label{Eq2}
\Delta\equiv (r-M)^2\ \ \ \ ; \ \ \ \ \Sigma\equiv
r^2+M^2\cos^2\theta\  .
\end{equation}
The degenerate horizon of the extremal black-hole spacetime is
defined by the radial functional relation
$\Delta(r=r_{\text{H}})=0$, which yields the simple relation
\begin{equation}\label{Eq3}
r_{\text{H}}=M\  .
\end{equation}

The dimensionless energy ratio $E(r)/\mu$ (energy per unit mass as
measured by asymptotic observers) associated with a circular orbit
of radius $r$ in the equatorial plane of the extremal
(maximally-spinning) Kerr black-hole spacetime is given by the
functional expression \cite{tw1}
\begin{equation}\label{Eq4}
{{E(r)}\over{\mu}}={{{r\pm
M^{1/2}r^{1/2}-M}}\over{r^{3/4}(r^{1/2}\pm 2M^{1/2})^{1/2}}}\  .
\end{equation}
The upper/lower signs in the dimensionless energy expression
(\ref{Eq4}) correspond to co-rotating/counter-rotating circular
orbits in the black-hole spacetime, respectively. From Eq.
(\ref{Eq4}) one finds the near-horizon co-rotating energy ratio
\cite{Noteco}
\begin{equation}\label{Eq5}
{{E(x)}\over{\mu}}={{1}\over{\sqrt{3}}}+{{2}\over{3\sqrt{3}}}\cdot
x+O(x^2)\  ,
\end{equation}
where the dimensionless parameter $x$ (with the near-horizon
property $x\ll1$) is defined by the relation
\begin{equation}\label{Eq6}
r\equiv M(1+x)\  .
\end{equation}

\section{The Thorne hoop conjecture and a lower bound on the radii
of circular orbits in the extremal black-hole spacetime}

The total energy (as measured by asymptotic observers) of the
composed extremal-Kerr-black-hole-orbiting-particle system is given
by
\begin{equation}\label{Eq7}
E_{\text{total}}=M+E_{\text{ISCO}}+O(E^2_{\text{ISCO}}/M)\  .
\end{equation}
According to the Thorne hoop conjecture \cite{Thorne}, the
composed system will form an engulfing horizon if it can be placed inside a ring
whose circumference $C$ is equal to (or smaller than) $4\pi E_{\text{total}}$.
That is, Thorne's famous hoop conjecture asserts that
\cite{Thorne}
\begin{equation}\label{Eq8}
C(E_{\text{total}})\leq 4\pi E_{\text{total}}\ \  \Longrightarrow \
\ \text{Black-hole horizon exists}\ .
\end{equation}

As we shall now show, the hoop relation (\ref{Eq8}) provides
compelling evidence for the existence of a lower bound (with
$r_{\text{circular}}>r_{\text{min}}>r_{\text{H}}$) on the radii of
circular orbits in the maximally-spinning (extremal) Kerr black-hole
spacetime. In particular, using Eqs. (\ref{Eq1}) and (\ref{Eq2}) one
obtains the functional expression \cite{Notertt0,Notenqq}
\begin{equation}\label{Eq9}
C(r)=2\pi\sqrt{r^2+M^2+2M^3/r}\  ,
\end{equation}
for the circumference $C=C(r)$ of an equatorial circular orbit in
the extremal Kerr black-hole spacetime. Substituting (\ref{Eq6})
into (\ref{Eq9}), one finds the simple near-horizon relation
\begin{equation}\label{Eq10}
C(x\ll1)=4\pi
M\cdot\big[1+{{3}\over{8}}\cdot x^2+O(x^3)]\  .
\end{equation}
Taking cognizance of Eqs. (\ref{Eq5}), (\ref{Eq7}), (\ref{Eq8}), and
(\ref{Eq10}), one obtains the dimensionless lower bound
\begin{equation}\label{Eq11}
x_{\text{circular}}>x_{\text{min}}=\Big({{8}\over{3\sqrt{3}}}\cdot
{{\mu}\over{M}}\Big)^{1/2}\  .
\end{equation}
on the scaled radii of circular orbits in the composed
extremal-Kerr-black-hole-orbiting-particle system. In particular,
according to the Thorne hoop conjecture \cite{Thorne}, composed
black-hole-particle configurations whose circular orbits are
characterized by the relation $x_{\text{circular}}\leq
x_{\text{min}}$ are expected to be engulfed by a larger horizon with
$r_{\text{horizon}}\geq r_{\text{circular}}$ [see Eqs. (\ref{Eq8})
and (\ref{Eq11})].

\section{Summary}

A remarkable feature of the maximally-spinning (extremal) Kerr
black-hole spacetime, which has been discussed in the physics
literature by many authors (see
\cite{tw1,tw2,tw3,tw4,Ted1,Ted2,Ted3} and references therein), is
the existence of co-rotating circular orbits which are characterized
by the limiting radial behavior
$r_{\text{circular}}/r_{\text{H}}\to1^+$.

In the present compact paper we have used the famous Thorne hoop
conjecture \cite{Thorne} in order to provide evidence for the
possible existence of a larger horizon (with $r_{\text{horizon}}\geq
r_{\text{circular}}\geq r_{\text{H}}$) that engulfs composed
extremal-Kerr-black-hole-orbiting-particle configurations which
violate the dimensionless relation (\ref{Eq11}). In particular, our
analysis has revealed the intriguing fact that circular orbits in
the extremal Kerr black-hole spacetime are restricted to the radial
region [see Eqs. (\ref{Eq3}), (\ref{Eq6}), and (\ref{Eq11})]
\begin{equation}\label{Eq12}
r_{\text{circular}}>r_{\text{H}}\cdot\Big[1+\Big({{8}\over{3\sqrt{3}}}\cdot
{{\mu}\over{M}}\Big)^{1/2}+O\Big({{\mu}\over{M}}\Big)\Big]\  .
\end{equation}

\bigskip
\noindent {\bf ACKNOWLEDGMENTS}
%\bigskip

This research is supported by the Carmel Science Foundation. I would
like to thank Yael Oren, Arbel M. Ongo, Ayelet B. Lata, and Alona B.
Tea for stimulating discussions.


\begin{thebibliography}{99}

\bibitem{tw1} J. M. Bardeen, W. H. Press and S. A. Teukolsky, Astrophys. J. {\bf 178}, 347 (1972).

\bibitem{tw2} S. Chandrasekhar, {\it The Mathematical Theory of Black Holes}, (Oxford
University Press, New York, 1983).

\bibitem{tw3} S. L. Shapiro and S. A. Teukolsky, {\it Black holes, white dwarfs, and
neutron stars: The physics of compact objects} (Wiley, New York,
1983).

\bibitem{tw4} S. Hod, Phys. Rev. D {\bf 84}, 104024 (2011) [arXiv:1201.0068]; S.
Hod, Phys. Rev. D {\bf 84}, 124030 (2011) [arXiv:1112.3286]; S. Hod,
Phys. Lett. B {\bf 718}, 1552 (2013) [arXiv:1210.2486].

\bibitem{Ted1} T. Jacobson, Class. Quant. Grav. {\bf 28}, 187001
(2011).

\bibitem{Ted2} P. P. Pradhan and P. Majumdar, The Euro. Phys. Jour. C {\bf 73}, 2470 (2013).

\bibitem{Ted3} S. Ulbricht and R. Meinel, Class. Quant. Grav. {\bf 32}, 147001
(2015).

\bibitem{Noteext} Note that maximally spinning Kerr black-hole spacetimes are
characterized by the relations $M=a=r_{\text{H}}$, where
$\{M,a,r_{\text{H}}\}$ are respectively the mass, angular momentum
per unit mass, and the degenerate horizon radius of the extremal
black hole.

\bibitem{Noteunit} We shall use natural units in which $G=c=1$.

\bibitem{Thorne} K. S. Thorne, in {\it Magic without Magic: John Archibald Wheeler},
edited by J. Klauder (Freeman, San Francisco, 1972).

\bibitem{Notemu} Here $\mu$ is the proper mass (with $\mu\ll M$) of the orbiting
particle.

\bibitem{Noteco} We shall henceforth focus our attention on the
{\it co}-rotating circular orbits of the Kerr black-hole spacetime.
These are the orbits which approach the black-hole horizon
($r^{\text{co-rotating}}_{\text{circular}}/r_{\text{H}}\to 1^+$) in
the extremal $a/M\to1^-$ limit.

\bibitem{Notertt0} Here we have substituted the characteristic
relations $dt=dr=d\theta=0$, $\theta=\pi/2$, and $\Delta\phi=2\pi$
for equatorial circular orbits in the line element (\ref{Eq1}) of
the extremal Kerr black-hole spacetime.

\bibitem{Notenqq} It is worth noting that the circumference of an engulfing hoop
which is perpendicular to the equatorial plane (that is, with the
properties $dt=dr=d\phi=0$ and $\Delta\theta=2\pi$) of the
maximally-spinning (extremal) black hole is smaller than the
calculated circumference (\ref{Eq9}) which characterizes the
equatorial $(\theta=\pi/2)$ engulfing hoop [this simple fact stems
from Eqs. (\ref{Eq1}) and (\ref{Eq2}) with the characteristic
property $\Sigma<r^2+M^2+2M^3/r$]. One therefore concludes that if
the equatorial hoop (\ref{Eq9}) is characterized by the inequality
(\ref{Eq8}), then the perpendicular engulfing hoop will also be
characterized by the same inequality.

\end{thebibliography}
\end{document}